\documentclass[
 reprint,
 amsmath,amssymb,
 aps,
 pra
]{revtex4-2}

\usepackage{newtxtext,newtxmath}

\usepackage{multirow}
\usepackage{graphicx}
\usepackage{bm}

\usepackage{subcaption}
\usepackage{tikz}
\usepackage[colorlinks=true]{hyperref}
\usepackage{xcolor}
\hypersetup{
  linkcolor=blue,      
  citecolor=blue!60!black, 
  urlcolor=magenta     
}
 
\usepackage{caption}
\begin{document}
\title{Extended three dimensional optical tweezers with a Single Gaussian Beam in a stratified medium}


\author{Sramana Das}
\author{Suvajit Dey}
\author{Nirmalya Ghosh}
\author{Subhasish Dutta Gupta}
\altaffiliation{Tata Institute of Fundamental Research, Hyderabad, Telangana 500046, India}
\author{Ayan Banerjee}
\affiliation{Department of Physical Sciences, Indian Institute of Science Education and Research Kolkata, Mohanpur, India- 741246}

\email{ayan@iiserkol.ac.in}

\begin{abstract}
We demonstrate a simple and effective approach for realizing extended volumetric optical trapping using a single linearly polarized Gaussian beam propagating through a stratified medium with engineered refractive index gradients. On modifying the gradients, the focal field undergoes axial elongation and develops multiple localized intensity maxima, enabling simultaneous trapping of particles at distinct axial planes. Both on-axis and off-axis trapping configurations are experimentally observed, forming a volumetric particle distribution without the need for complex holographic beam shaping. A rigorous theoretical framework based on the Debye–Wolf diffraction formalism combined with Generalized Lorenz–Mie Theory is employed to model the focused field and the resulting optical forces. The analysis, substantiated with numerical simulations, shows that spherical aberration arising from refractive-index discontinuities plays a crucial role in redistributing optical energy, leading to the formation of multiple stable trapping regions. The axial trapping potential decreases monotonically with increasing mismatch, indicating a trade-off between multi-site trapping capability and trapping potential, which is in quantitative agreement with experimental observations. This work establishes refractive index engineering as a powerful tool for tailoring three-dimensional optical force landscapes using minimal optical complexity, thus offering a scalable alternative to conventional holographic optical tweezers towards applications in colloidal assembly, micro-manipulation, and biophotonics.
\end{abstract}

\keywords{Optical Tweezers, Optical Force, Tight focusing, Trapping of micro-particles, Stratified medium.}

 \maketitle

\section{Introduction}
\begin{figure*}[t]
    \centering
    \includegraphics[width=1\linewidth]{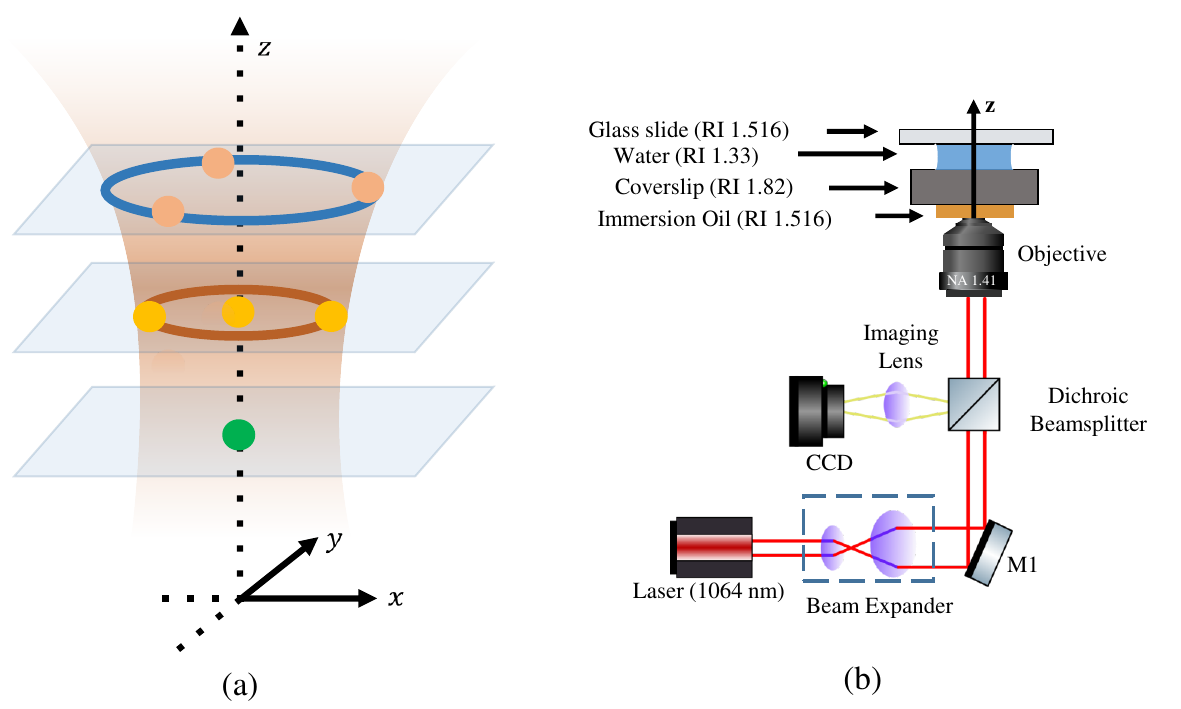}
    \caption{(a) Gaussian Beam trajectory showing particles trapped in different axial planes (b) Schematic representation of the experimental setup. }
    \label{fig:schematic}
\end{figure*}
Optical tweezers harness the highly localized intensity gradients of tightly focused laser beams to exert optical forces capable of trapping and manipulating mesoscopic particles with remarkable precision \cite{Ashkin1986}. Since their development, they have become an indispensable tool across physics, chemistry, and the life sciences \cite{Pesce2020,Marago2013}, owing to their unique ability to achieve stable confinement of cells, colloids, and nanoparticles in the presence of stochastic Brownian forces, thereby facilitating detailed investigation of mechanical and dynamical processes \cite{Neuman2008,Bustamante2021}. Their non-invasive character and natural compatibility with optical microscopy have enabled a broad spectrum of applications, ranging from single-molecule force spectroscopy and mechanobiology to colloidal self-assembly and micro-engineering \cite{Neuman2008,Bustamante2003}. While this versatility has driven tremendous progress in single-particle studies \cite{JonesOpticalTweezers2015}, a growing class of problems in contemporary science demands something more: the simultaneous control of multiple particles to probe collective interactions, emergent group behavior, and the formation of complex three-dimensional assemblies \cite{Grier2003}. Meeting this demand requires stable, independent confinement of multiple particles at spatially distinct locations in all three dimensions, which therefore remains one of the most persistent and fundamental challenges in the field.

Towards this, considerable effort has gone into extending optical trapping from a single focal spot to multiple simultaneous sites \cite{Grier2003}. Holographic optical tweezers based on spatial light modulators (SLMs) represent the most flexible of these approaches, generating reconfigurable trap arrays capable of confining tens to hundreds of particles in arbitrary spatial arrangements \cite{Li2021}, which are however principally in the direction transverse to the propagating beam. For atoms, however, 3D structures have been routinely achieved by electric field structures generated by SLMs due to the extremely small confinement volumes required for that purpose \cite{Lee2016}. For larger particles, generating a strong enough electric field in the axial direction over extended distances remain a challenge. To address this issue, structured beams including vortex modes and higher-order Laguerre–Gaussian beams exploiting engineered phase and intensity profiles to create multiple intensity peaks \cite{Padgett2011} have been deployed -- with diffractionless beams such as Bessel and Airy beams, which offer an extended depth of field and multiple high-intensity lobes along the propagation direction \cite{Durnin1987}, being of particular interest. However, even for such beams, the intensity contrast between these lobes is often very high, making it difficult to sustain robust, independent trapping of separate particles axially. Diffractive optical elements and multi-core fiber configurations have also been employed towards similar ends, but these are again mostly designed for trapping multiple particles in the radial or transverse direction \cite{Cizmar2011NatPhoton, Curtis2002}. Optical traps employing counterpropagating or retroreflected laser beams have led to the trapping of multiple Rayleigh particles axially \cite{Constable1993}, but these on the other hand, suffer from the inability of confining multiple particles in the radial direction in a stable manner. In addition, almost all the above-mentioned approaches rely on external beam-shaping hardware or multiple lasers that add cost, complexity, and practical constraints to the optical system. An elegant solution capable of achieving stable three-dimensional multi-particle trapping using a single, fundamental Gaussian beam without augmentation by auxiliary optical elements has still remained out of reach. It is precisely this gap in the literature that motivates the work that we present in this paper. Beyond its conceptual appeal, such a capability would open direct experimental access to the study of many-body hydrodynamic interactions in colloidal suspensions, near-field coupling between trapped mesoscopic particles, and the dynamic assembly of three-dimensional microstructures in soft condensed matter systems \cite{Grier2003}.

Thus, in this paper, we introduce a fundamentally unique strategy that achieves stable multi-particle trapping in three dimensions near the focal region of an optical trap using nothing more than a single tightly focused Gaussian beam. Our approach exploits spherical aberration arising from a deliberate refractive index mismatch between the immersion oil and the coverslip (thereby creating a refractive index stratified medium) to engineer distinct, well-separated high-intensity lobes along and perpendicular to the trapping beam's propagation axis. Rather than treating spherical aberration as an imperfection to be corrected, we harness it as a controllable physical mechanism for sculpting the three-dimensional intensity profile of the focused beam \cite{Torok:97,vandeNes2004Vectorial}. Experimentally, we demonstrate the trapping of colloidal particles (diameter corresponding to the Mie regime) at different axial depths both at the centre of the beam, and radially outwards for different refractive index (RI) contrasts of the stratified medium. To explain our experimental observations and characterize the optical landscape of our trapping system, we develop a theoretical framework based on the Generalized Lorenz–Mie theory extended to stratified media \cite{neves2019optical,bohren_huffman_1983,Dutra2016PRA}. This enables us to compute the three-dimensional optical potential experienced by a trapped particle, which we match with our experimental measurements with reasonable accuracy. Importantly, given the appropriate input parameters, our methodology allows the possibility to calculate optical forces experienced by particles in the Mie regime trapped optically, thereby opening up multiple applications in force spectroscopy using optical tweezers \cite{JonesOpticalTweezers2015}. Our approach also eliminates the need for SLMs, diffractive elements, or wavefront shaping, offering a robust, low-cost, and easily implementable platform for three-dimensional multi-particle manipulation compatible with standard microscopy infrastructure.

 \section{Experiment}
In our earlier work, we showed the trapping and manipulation of colloidal particles at axial distances of around 2 $\mu$m from the focus using a RI stratified medium, and observed very interesting effects demonstrating the spin-orbit interaction of light, viz. the spin-Hall effect \cite{roy2014manifestations, das2025comprehensive}, transverse spin \cite{kumar2024probing}, etc. Thus, while trapping at distances away from the focus seemed possible using such stratified media, the role of the latter in simultaneous trapping at different axial and radial planes had not yet been explored. Our first task was thereby to set up an experiment where the RI stratified medium could be varied, and their effects on the trapping of colloidal particles near the focal plane could be studied carefully, and whether configurations such as those depicted in the cartoon in Fig.~\ref{fig:schematic}(a) could be achieved. 

The schematic and details of our experimental setup is shown in Fig.~\ref{fig:schematic}. We used  a conventional optical tweezers configuration consisting of an inverted microscope (Carl Zeiss Axioert.A1) with an oil-immersion 100X objective (Zeiss, NA 1.4) and a laser of wavelength 1064 nm coupled to the back port of the microscope. We then focused the linearly polarized light into the sample chamber of a 4-layered stratified medium shown in Fig.\ref{fig:schematic}(b). The stratified medium consisted of (a) an oil layer
of thickness around $5\,\mu\mathrm{m}$ and refractive index (RI) 1.516, (b)
a $200\,\mu\mathrm{m}$-thick coverslip having refractive index 1.516/1.814
(henceforth referred to as “matched”/“mismatched” conditions, respectively; note that the matched condition is typically employed in optical tweezers to minimize spherical aberration effects at the focal spot) and (c) sample chamber (water solution of Polystyrene of average diameter $3\,\mu\mathrm{m}$) having a refractive index of 1.33 with a depth of $20 ~\mu\mathrm{m}$, and finally (d) a glass slide of refractive index 1.516 on the top.

\begin{figure}[t]
    \centering
    \includegraphics[width=1\linewidth]{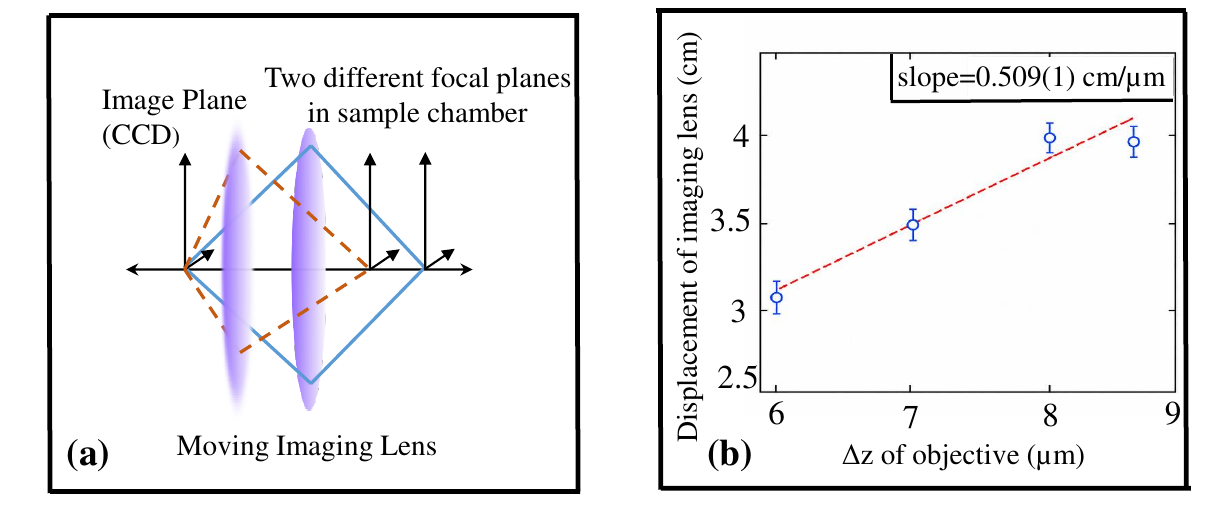}
    \caption{(a) Schematic depicting how the imaging lens is moved to capture two different z-planes. (b) Calibration of the imaging lens - the slope represents the ratio of the distance moved by the external lens in cm to obtain a clear image on the camera versus the movement of the objective lens in $\mu$m is around 0.509(1) cm/$u$m.}
    \label{calibration1}
\end{figure}


\begin{figure*}[t]
    \centering
    \includegraphics[width=1\linewidth]{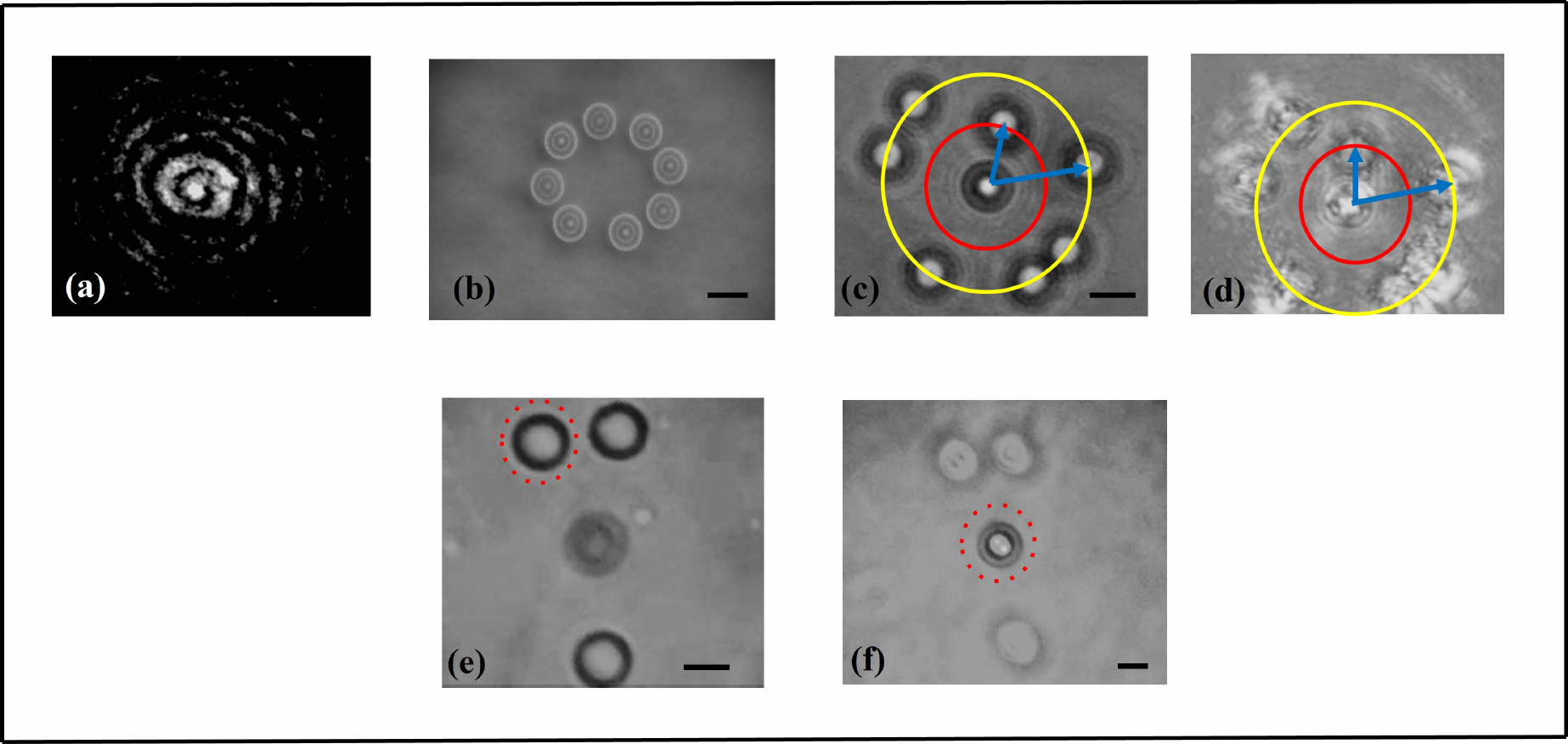}
    \caption{(a) CCD camera image of trapping intensity pattern taken in the back focal plane of the microscope while using the coverslip of refractive index 1.659. (b) Annular ring structure of polystyrene particles of diameter 2 $\mu$m formed using the coverslip of refractive index 1.659. (c) Particles trapped in secondary maxima of two different radii (image taken with a filter for 1064 nm) when a coverslip of refractive index 1.814 was  used. (d) Same as (c) without the filter showing the intensity pattern of the beam at trapping positions. (e) 2 $\mu$m diameter particles trapped at different axial positions -- by changing the imaging lens position, we first bring the particles trapped off-axis at higher $z$ in focus (shown by the dotted red circle, note that the particle at the center is blurred), and then (f) bring the particle in the center in focus (again shown by the dotted red circle, note that now the off-axis particles are blurred). All scale bars in the figures correspond to a distance of 3 $\mu$m.}
    \label{expFigure}
\end{figure*}

Since we attempted to trap particles at different axial planes inside the sample chamber, it was necessary to image the different axial planes on an external CCD camera. For this purpose, a convex lens with a focal length of 7.5 cm was placed in front of the camera, as shown in Fig.\ref{fig:schematic}(b). A stuck polystyrene bead adhered to the coverslip served as the calibration target. By translating the imaging lens back and forth  along the optical axis, different axial planes were brought into focus [Fig.\ref{calibration1}(a)]. The displacement of the imaging lens was then calibrated with respect to the axial movement of the objective lens [Fig.\ref{calibration1}(b)] and yielded a ratio(slope) of 0.509(1) cm/$\mu$m, allowing accurate determination of the axial separation between the trapping planes. In addition, we determined the trap corner frequencies employing the Power Spectral Density method \cite{neuman2004optical}, from which the optical trapping forces applied on the micro-particles could be estimated. Finally, we validated these experimental results with theoretical simulations provided in the next section.

\subsection{Observations}
We used coverslips of two different RIs - 1.659 and 1.814, resulting in distinct refractive index gradients in the medium, as well as two different objective lenses with numerical apertures (NAs) of 1.3 and 1.4. We first recorded the beam intensity pattern at the back focal plane of the microscope while using the coverslip of refractive index 1.659 and the objective lens of NA 1.4 as shown in Fig.\ref{expFigure}(a). The pattern showed was as expected for such RI stratified media -- with a central ring structure \cite{roy2013controlled}  surrounded by weak Airy rings. Note that this pattern results due to the  superposition of the intensity fields at all axial planes in the sample, which are made up of scattered fields from the microsphere and reflected fields from the different surfaces of the sample chamber. We then observed the spontaneous formation of annular assemblies of optically trapped microparticles, as illustrated in Fig.~\ref{expFigure}(b). The particles were allowed to accumulate progressively and, upon slight adjustment of the microscope focus, reorganized into the well-defined closed ring structures. Interestingly, particles could be simultaneously trapped at distinct secondary intensity maxima of varying radii, as demonstrated in Fig.~\ref{expFigure}(c) and (d), which are taken with and without a 1064 nm filter placed in front of the camera, respectively. Thus, particle assemblies at different radial positions for the same $z$ (axial position) can be achieved, akin to two-dimensional holographic tweezers. 

We now move on to the demonstration of the central result of this study, which substantiates the holographic tweezers scheme generated by our design. Figures~\ref{expFigure}(e) and (f) (Video S1 and S2) demonstrate the simultaneous realization of radially off-axis and on-axis trapping at two distinct axial planes respectively. While particles are trapped at different spatial locations, we are able to bring the particles trapped off-axis into sharp focus by changing the position of our external imaging lens as shown in Fig.~\ref{expFigure}(e), followed by the particle at the center in Fig.~\ref{expFigure}(f). On both occasions, the other particles, i.e. the one in the center in Fig.~\ref{expFigure}(e), and the ones off-axis in Fig.~\ref{expFigure}(f), appear blurred. 

\begin{figure*}[t]
    \centering
    \includegraphics[width=1\linewidth]{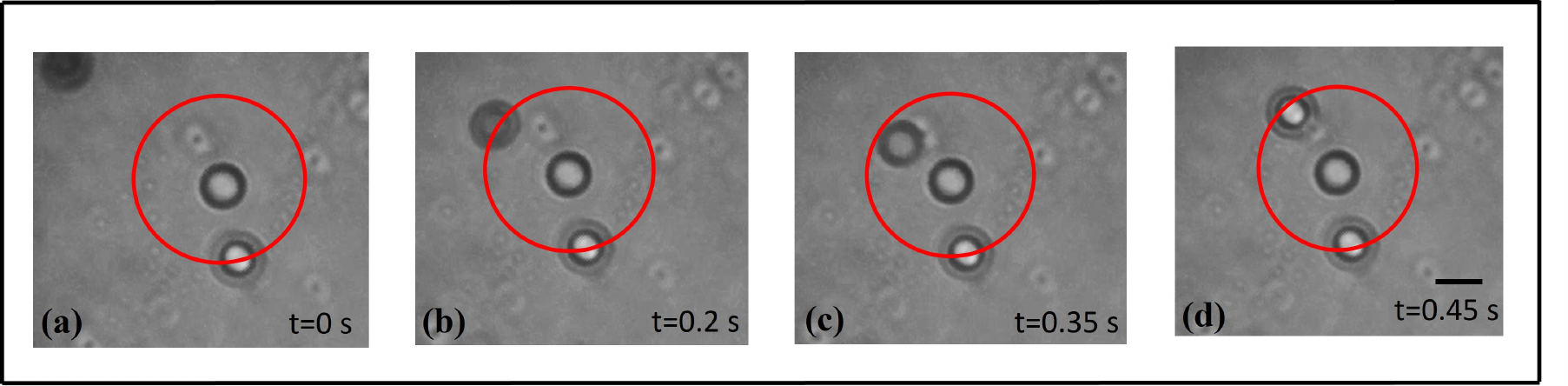}
    \caption{ Time-lapsed frames from Video 1, depicting a trapped particle at the trap focus (centre of circle drawn in red solid line to depict the radial distance from the central particle) and also sharply focused on the camera, along with a particle trapped off-axis below the central particle (at the right lower edge of the circle) at a different $z$-plane, as is clear from the image sharpness which is different from the central particle. (a) shows a third particle at the top left corner, at a different $z$-plane from the other two, as apparent from its unfocused image. (b) the particle is drawn towards the trap and is at the left upper edge of the red circle, and is still unfocused. (c) the particle comes closer to the central particle, and is now within the red circle. (d) the particle is seen repelled back to the left upper edge of the red circle, and is now trapped off-axially. The $z$-plane is clearly similar to the other particle trapped off-axis at the left bottom of the circle. The scale bar corresponds to 3 $\mu$m.}
    \label{fig:timesnap}
\end{figure*}

Furthermore, in a different set of experiments depicted in Video S3 (and time lapse images from the video in Fig.~\ref{fig:timesnap}), we first observe a particle trapped at the center (sharp focus) and another radially separated and at a different $z$-plane (slightly out of focus), as shown in Fig.~\ref{fig:timesnap}(a). A particle approaches from the top left in Fig.~\ref{fig:timesnap}(b) and appears very much blurred, so that it clearly traverses a different $z$-plane compared to both the trapped particles. In Fig.~\ref{fig:timesnap}(c), this particle is seen to approach very close to the central particle both radially and axially as is understandable from the reduced radial distance, and the sharper focus, respectively. However, in Fig.~\ref{fig:timesnap}(d), the particle is seen to withdraw farther radially from the central particle, and is then stably trapped. This suggests the existence of a net repulsive force at a particular location in the radial direction (observed clearly in Video S2) which is the result of the focused and scattered electric fields, that we investigate in the following sections by numerical simulations of our system.

\begin{figure*}[t]
    \centering
    \includegraphics[width=1\linewidth]{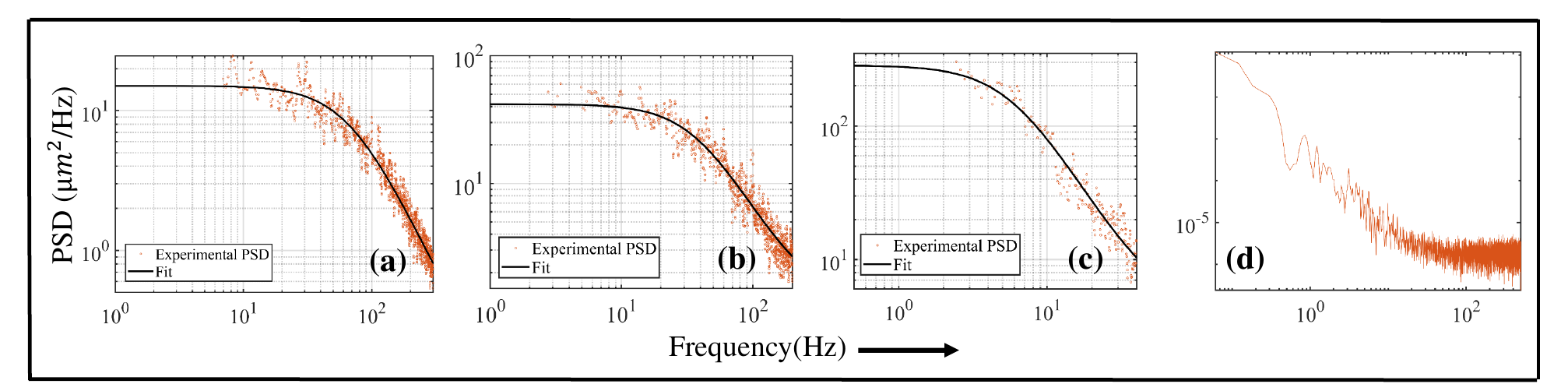}
    \caption{ Power Spectrum Density plots to determine the corner frequency for (a) “matched” case (b) “mismatched” on-axis (c) “mismatched” off-axis (d) free particle }
    \label{fig:psd}
\end{figure*}

To experimentally quantify the optical forces, we measured the power spectral density (PSD) of trapped particles for both “matched” and “mismatched” configurations, as shown in Fig.~\ref{fig:psd}(a) - (c). For comparison, the PSD of a freely diffusing particle is also presented in Fig.~\ref{fig:psd}(d). Time-series data were acquired by recording 30 s videos of the trapped particles for both configuration using a fast camera. The PSDs were subsequently plotted using MATLAB and IGOR Pro, fitted with a Lorentzian function, and the corresponding corner frequencies were extracted for each case.

 \subsection{Analysis}
The experimental observations described above are further analyzed to quantitatively characterize the trapping dynamics, underlying optical forces, and the extended three-dimensional trapping configuration achieved using our system. To complement the qualitative results and gain deeper insight into the observed behavior, we performed a detailed analysis of particle motion using power spectral density (PSD) measurements. This approach enables us to evaluate the strength and nature of the trapping potential, thereby elucidating the trapping characteristics and validating the extended trapping volume facilitated by our system.

As we described earlier, we use an external imaging system to determine the physical separation of the different axial planes in which particles are trapped. We now discuss the calibration methodology in detail. As shown in Fig.~\ref{calibration1}(a), controlled translation of the imaging lens positioned in front of the CCD enables selective focusing of different axial planes, thereby allowing particles at distinct depths to be independently resolved.
The corresponding quantitative analysis, shown in Fig.~\ref{calibration1}(b), yields a mean value of 0.509(1) for the calibration factor of converting the external measurement to the actual physical separation in the sample chamber, where the value in parentheses denotes the uncertainty or error in the final digit. This result is obtained from multiple experimental measurements and demonstrates good consistency across repeated trials. The small value of the measurement uncertainty confirms the reliability of the calibration procedure and supports the existence of well-defined, spatially separated trapping planes, with the separation coming out to be around 2(0.1) $\mu$m. Thus, the particle trapped at the center in Fig.~\ref{expFigure}(f) appears to be at the beam focus, whereas those trapped radially off-axis are estimated at a plane 2(0.1) $\mu$m further[Fig.~\ref{expFigure}(e)]. This provides definitive evidence of the spatially extended three dimensional nature of the optical field, which is able to support optical trapping over a large spatial volume, unprecedented in the existing literature to the best of our knowledge. 
The trapped particle ring [\ref{expFigure}(a)] exhibits a transverse (radial) extent of approximately 5 $\mu$m from the trap center. This experimentally observed value was subsequently validated through theoretical modeling.
\begin{table}[h]
\centering
\renewcommand{\arraystretch}{1.5}
\begin{tabular}{|c|c|c|c|c|c|}
\hline
\textbf{RI Stratification} &
\textbf{Case} &
$(f_c)_x$ &
$(f_c)_y$ &
$(f_c)_{\rm avg,x}$ &
$(f_c)_{\rm avg,y}$ \\
\hline

Matched & 1 & $71 \pm 10$ & $74 \pm 9$ &  & \\ \hline
        & 2 & $84 \pm 9$& $74 \pm 10$& $74.6  \pm 8.3$& $71.6 \pm 8.6$ \\ \hline
        & 3 & $69 \pm 6$ & $67 \pm 7$ & & \\ \hline

Mismatch (On-Axis) & 1 & $39 \pm 3$& $48 \pm 4$& & \\       \hline
                 & 2 &$41 \pm 4$ &$37 \pm 2$ &$39.6 \pm 3.3$ & $40.3 \pm 2.3$\\ \hline
                 & 3 & $39 \pm 3$& $36 \pm 1 $& & \\ \hline

Mismatch (Off-Axis) & 1 &$6.3 \pm 1$ & $7.2 \pm 2$& & \\ \hline
                  & 2 &$7.3 \pm 2$ & $6.8 \pm 1.6$& $7.7 \pm 2$& $6.6 \pm 1.7$\\ \hline
                  & 3 & $9.5 \pm 3$&$6 \pm 1.6$ & & \\ \hline

\end{tabular}
\caption{Corner Frequencies Along the x- and y-Directions for Different RI Stratification Cases}
\label{tab:RI_cases}
\end{table}
We now proceed to provide a quantitative estimate of the optical forces derived from the experimental results. The trapping stiffness and the corresponding optical forces were quantitatively evaluated by power spectral density (PSD) of particle position fluctuations [Fig.~\ref{fig:psd}]. The experimentally obtained PSDs were fitted with a Lorentzian function, from which the corner frequency ($f_{c}$) was extracted for each trapping condition, as shown in Table \ref{tab:RI_cases}. Since the corner frequency is directly proportional to the stiffness of the trap, variations in $f_{c}$ in different cases reflect changes in the strength of the optical confinement. As shown in Table 1, a clear distinction in the corner frequency values was observed between the matched and mismatched configurations, indicating a significant modification of the trapping potential due to the refractive index contrast. In contrast, the PSD of a freely diffusing particle does not exhibit a characteristic corner frequency, confirming the absence of confinement [Fig.~\ref{fig:psd}(d)].

A quantitative comparison of the extracted corner frequencies ($f_{c}$) reveals a substantial variation between on-axis (On) and off-axis (Off) trapping under matched and mismatched configurations. For the mismatched case at the on-axis position, the corner frequencies are reduced to 52\% and 55\% of the corresponding matched values along the x- and y-directions, respectively. On the other hand, for off-axis trapping for the mismatched case, a significantly stronger reduction is observed. The corner frequencies decrease to 19\% (x-direction) and 16\% (y-direction) relative to the mismatched on-axis values. When directly compared to the matched configuration, the off-axis corner frequencies further reduce to approximately 10\% and 9\% of the matched values along the x- and y-directions, respectively. Overall, the data indicate that the corner frequency -- and hence the trapping strength -- are highest in the matched on-axis condition, decrease under mismatched-on-axis trapping, and are further reduced in the off-axis regime which we later validated . 

We now attempt to understand our experimental observations through numerical simulations of the system. 

\section{Theoretical Analysis}
In our simulations, we specifically validate the following  results we obtained in experiments: (i) the axial separation between the trapping planes, confirming the volumetric nature of the trapping potential, (ii) the transverse extent of the trapping region, and (iii) the trapping strength from the depth of the optical potential. We first lay down our theoretical formalism.

To investigate the optical trapping characteristics of the system, we employ the Generalized Lorenz–Mie Theory (GLMT) combined with the Debye–Wolf diffraction formalism \cite{neves2019optical, PhysRevE.76.061917}. This framework provides a rigorous description of the electromagnetic field generated by a tightly focused Gaussian beam propagating through a stratified medium and its interaction with a dielectric particle. Within the Debye–Wolf representation, the non-paraxial field emerging from a high–numerical-aperture objective is expressed as a continuous superposition of plane waves. The electromagnetic field is decomposed into transverse electric ($s$) and transverse magnetic ($p$) spatial harmonics, allowing the amplitude and phase modifications introduced by each dielectric interface to be incorporated through the corresponding Fresnel transmission coefficients.

\begin{figure*}[t]
    \centering
    \includegraphics[width=1\linewidth]{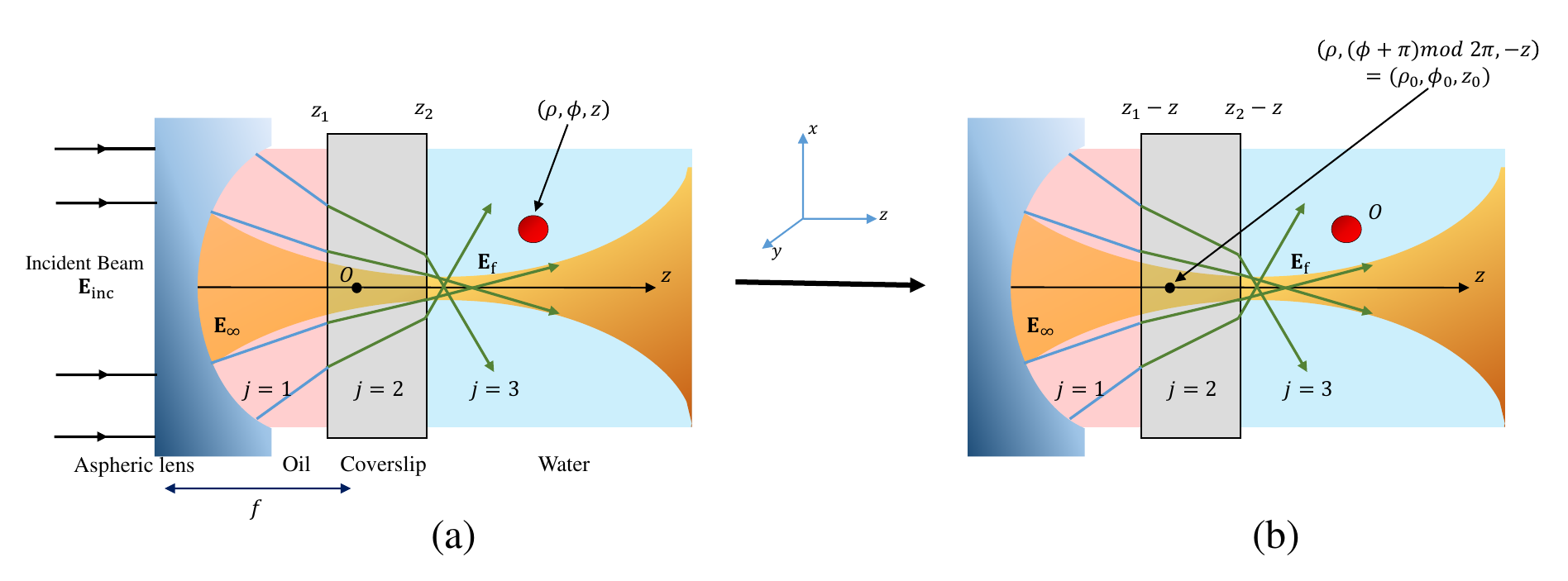}
    \caption{Schematic representation of the coordinate transformation used in the GLMT formulation. (a) Laboratory frame, where the incident field $\mathbf{E}_{\mathrm{inc}}$ is focused through the stratified oil--coverslip--water system and the particle is located at $(\rho_0,\phi_0,z_0)$. (b) Particle-centered frame obtained by translating the coordinate origin to the microsphere centre, with the interface positions shifted to $z_1-z_0$ and $z_2-z_0$.} \label{fig:coordinate}
\end{figure*}
In the experiment -- as mentioned earlier -- the optical beam propagates through four dielectric layers above the objective (immersion oil, coverslip, water, and the top coverslip of the sample chamber). In the theoretical model, however, we consider a reduced three-layer system consisting of immersion oil, coverslip, and water. This approximation is justified because the particle primarily interacts with the transmitted field forming the optical trap in the vicinity of the focal region. The upper water–glass interface mainly introduces weak reflections that generate low-contrast axial interference fringes within the sample chamber. Such interference primarily modulates the longitudinal intensity distribution along the beam propagation direction and therefore affects predominantly the axial force component.In contrast, the transverse trapping force is governed by the local lateral intensity gradient near the focal region, where \(I\) denotes the local optical intensity, determining the gradient force \(\mathbf{F}\propto\nabla I\) responsible for radial confinement, where $I$ is the intensity of light \cite{Ashkin1986, neuman2004optical}. Since the reflected field contributes weakly to the transverse phase symmetry and the lateral intensity gradients responsible for radial trapping stability, its influence on the radial force distribution is negligible. Note also that in the experiments, we measure only the trap stiffness in the radial direction, so the contribution of the axial field itself is not determined in our measurements. Consequently, the three-layer stratified model captures the dominant physics governing the transverse trapping landscape while significantly simplifying the numerical implementation. As we shall observe later, the good agreement between the theoretical predictions and the experimentally observed trapping behavior further validates this approximation.

Fig.\ref{fig:coordinate} shows a schematic of the optical system and the associated coordinate frame. The $z$-axis is taken along the optical axis of the objective, with the origin ($z=0$) located at the focal plane corresponding to the focal length $f$. The dielectric interfaces are arranged sequentially along the axial direction. All media involved in the system are assumed to be linear, homogeneous, isotropic, and non-conducting. Since the trapped particles are positioned at distances of several wavelengths from the interfaces, the contribution of evanescent waves is neglected. As our experiment uses only the Gaussian beam, we will only consider a beam profile before focusing, which has a Gaussian spatial structure and general polarization as follows, 
\begin{align}
    \textbf{E}_{inc} = E_0 e^{-f^2 \sin^2 \theta/\omega^2}\begin{pmatrix}
p_x \\
p_y
\end{pmatrix}
\end{align}
 where $\omega$ is the beam waist of the incident paraxial beam before entering the objective, while $p_x$ and $p_y$ denote the complex amplitudes of the electric-field polarization constituents along the $x$- and $y$-directions, respectively. For a normalized polarization state, they satisfy $
    |p_x|^2+|p_y|^2=1.
$ With respect to the field on the exit pupil $\mathbf{E}_{\infty}(\theta, \phi)$, the resulting transmitted electric field in the vicinity of the focus at the $j$-th medium (we here are interested about water medium, $j=3$), is expressed in cylindrical coordinates $(\rho, \phi, z)$ as follows:

\begin{align}
\mathbf{E}_{\text{f}}(\rho, \varphi, z)
= \frac{ik_j f \, e^{ik_1 f}}{2\pi}
\int_0^{\theta_{\text{max}}}
\int_0^{2\pi}
\mathbf{E}_{\infty}(\theta, \phi) & \notag \\  \,
e^{ + i k_j z \cos\theta_j }\,
e^{ i k_1 \rho \sin\theta_1 \cos(\phi-\varphi) }\,
\sin\theta_1 \, d\theta \, d\phi &.
\label{eq:RWE}
\end{align}

where $\theta_j =\sin^{-1}(n_1\sin\theta_1/n_j),$ and the angular integration limit $\theta_{max}$, is determined by the numerical aperture of the objective lens and corresponds to the maximum acceptance angle, further constrained by the onset of total internal reflection arising from the stratified medium. By applying the boundary conditions at the interfaces, one obtains explicit expressions for the transmitted far-field components defined on the reference sphere \cite{Novotny_Hecht_2012}.
\begin{align}
    \mathbf{E}_\infty (\theta, \phi) =& \sqrt{\frac{n_1}{n_a}\cos \theta}\; \frac{\sin \theta_j}{\sin \theta_1} \times\notag \\ &( \;T_s[\mathbf{E}_{inc}\cdot \hat{\varphi}] \hat{\varphi} + T_p\;[\mathbf{E}_{inc}\cdot \hat{\rho}] \hat{\theta}_j \;) \label{eq:e_inf_t} 
\end{align}
where $n_a$ and $n_1$ are the refractive indices of the medium before and just after the focusing lens, respectively. The effective transmission and reflection coefficients $T_s$ and $R_s$ in medium $j$, respectively, have been calculated using the method described in \cite{vandeNes2004Vectorial} based on the refractive indices of mediums and position of interfaces. 


Since the magnetic field is also an integral part of the GLMT, we use Maxwell’s equations to write $\mathbf{H}_\infty = \frac{1}{Z} (\hat{k} \times \mathbf{E}_\infty) $ where $\hat{k} = \sin \theta_j \cos \phi \;\hat{x} + \sin \theta_j \sin \phi \; \hat{y} + \cos \theta_j \; \hat{z}$ and obtain the magnetic field at focal region. 

Applying the GLMT \cite{neves2019optical, PhysRevE.76.061917}, the beam-shape coefficients (BSCs), $G_{nm}^{\mathrm{TE}}$ and $G_{nm}^{\mathrm{TM}}$, of the total focal field, $\mathbf{E}_f$, are first determined to obtain the optical forces on the trapped particle. Here, the coefficients determine the relative weights of the transverse electric (TE) and transverse magnetic (TM) spherical modes, respectively, for integers $n \geq 1$ and $-n \leq m \leq n$. BSCs are computed with respect to the center of the microsphere, avoiding the truncation error associated with the translation theorem. As schematically illustrated in Fig.~\ref{fig:coordinate}(a), the incident field $\mathbf{E}_{\mathrm{inc}}$ is focused by an aspheric lens through the stratified oil--coverslip--water system, where the interfaces are located at $z_1$ and $z_2$, and the particle is positioned at $(\rho_0,\phi_0,z_0)$ in the laboratory coordinate system $(\rho,\phi,z)$. As shown in Fig.~\ref{fig:coordinate}(b), the coordinate origin is subsequently translated to the centre of the microsphere, such that the particle is located at the new origin and the interface positions become $z_1-z_0$ and $z_2-z_0$. The corresponding particle-centered coordinates are related to the laboratory-frame coordinates through $(\rho,(\phi+\pi)\bmod 2\pi,-z)=(\rho_0,\phi_0,z_0)$. Here, the regions $j=1$, $j=2$, and $j=3$ correspond to the oil, coverslip, and water media, respectively. Thus, $(\rho_0,\phi_0,z_0)$ specifies the nominal position of the beam focus relative to the chosen coordinate origin, while the particle-centered representation provides the appropriate coordinate system for evaluating the BSCs and subsequently calculating the optical forces.

Then, the focal field expressions are reformulated in spherical coordinates to determine the radial components of the fields. The reflection of the scattered fields from the planar interfaces is not considered in the present analysis. Then, the final form of BSCs of transmitted fields can be expressed as:  

\begin{widetext}
\begin{equation}
\begin{aligned}
G^{\mathrm{TM/TE}}_{nm}
&= -i^{\,n-m+1}\, k_j f \, e^{i k_1 f}\, e^{-i m \phi_0}
\sqrt{\frac{n_b}{n_a}}
\sqrt{\frac{4\pi(2n+1)}{n(n+1)}\frac{(n-m)!}{(n+m)!}}  \\
&\quad \times
\int_{0}^{\theta_{\mathrm{max}}}
\mathrm d\theta \, \sin\theta_1 \,
\sqrt{\cos\theta_1}\,
e^{-f^{2}\sin^{2}\theta_1/\omega^{2}}\,
e^{-i k_j z_0 \cos\theta_j} \\
&\quad \times
\Bigg(
\begin{pmatrix}
\dfrac{m\, J_m(k_1\rho_0\sin\theta_1)}{k_1\rho_0\sin\theta_1}\,
T_{s/p}\;\pi_n^{m}(\theta_j)
+ J_m'(k_1\rho_0\sin\theta_1) T_{p/s}\,
\tau_n^{m}(\theta_j)
\end{pmatrix}
\\
&\qquad\qquad
- i
\begin{pmatrix}
\dfrac{m\, J_m(k\rho_0\sin\theta)}{k_1\rho_0\sin\theta_1} T_{p/s}\,
\tau_n^{m}(\theta_j)
+ J_m'(k\rho_0\sin\theta) T_{s/p}\,
\pi_n^{m}(\theta_j)
\end{pmatrix}
\Bigg)
\\
&\quad \cdot
\begin{pmatrix}
\cos\phi_0 & \sin\phi_0 \\
-\sin\phi_0 & \cos\phi_0
\end{pmatrix}
\begin{pmatrix}
p_x , p_y \\
p_y , -p_x
\end{pmatrix}.
\end{aligned}
\label{eq:BSC_t}
\end{equation}
\end{widetext}

In the above expressions, $J_m$ denotes the Bessel function of the first kind
of order $m$, $P_n^{m}(\theta)$ are the associated Legendre functions,
$\pi_n^m(\theta) \equiv m\,P_n^{m}(\theta)/\sin\theta,$ 
and 
$\tau_n^m(\theta) \equiv \mathrm{d}P_n^{m}(\theta)/\mathrm{d}\theta.$ Finally, the time-averaged optical force $\mathbf{F}(\rho,\phi,z)$ on a dielectric particle positioned at any arbitrary position of the trap is calculated using the Maxwell stress tensor formalism following the expressions given in \cite{neves2019optical} with the help of Mie scattering coefficients \cite{bohren_huffman_1983}. 
\begin{figure*}[t]
    \centering
    \includegraphics[width=1\textwidth]{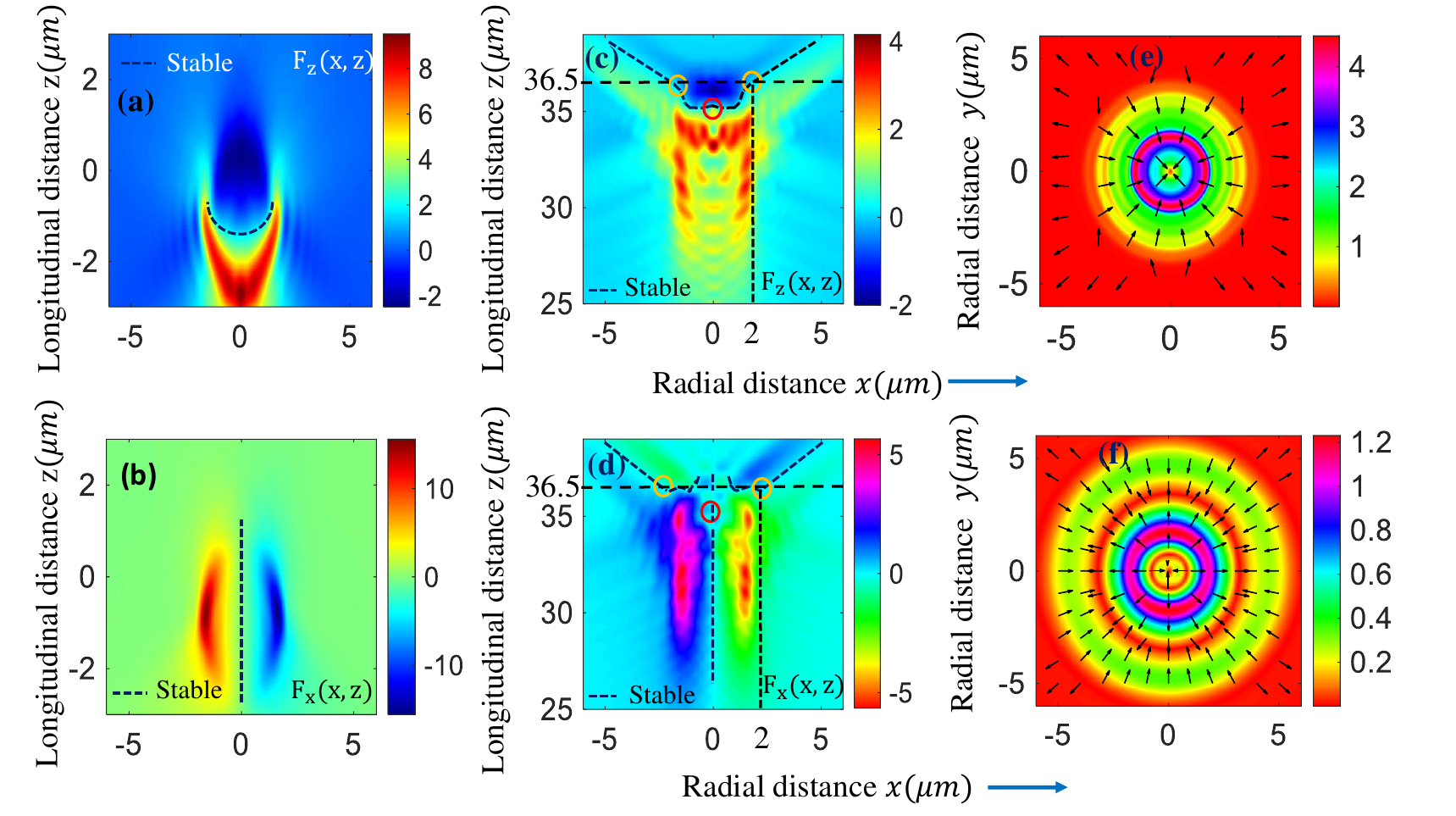}
   \caption{Optical force distributions under RI-matched and mismatched conditions. (a,b) Axial and radial optical force distributions in the longitudinal $xz$ plane ($y=0$) for the refractive-index matched case. A black dashed line shows the zero force contour for both $F_z$ (fig. (a)) and $F_x$ ((fig. (b)). (c,d) Corresponding axial and radial force distributions, respectively, for the refractive-index mismatched case. The zero force contours for both $F_z$ and $F_x$ have been substantially modified compared to the matched case. The red and golden circles represent the on-axis and off-axis trapping positions respectively, with the corresponding axial positions marked in the y-axis, respectively. (e,f) Heatmaps of the radial force magnitude, $|F_r|(x,y)=\sqrt{F_x^2+F_y^2}$, in the transverse $xy$ plane, overlaid with quiver arrows indicating the force direction, shown at two sample chamber heights: $z=35\,\mu\mathrm{m}$ (focal plane, (fig. (e))) and $z=37.2\,\mu\mathrm{m}$ (fig. (f)).}
    \label{fig:optical force}
\end{figure*}
\section{Numerical Simulations}
To gain further insight into the experimentally observed trapping behaviour, we analyse the system through numerical simulations implemented in MATLAB, following the methodology described in the previous section. In the simulations, a tightly focused laser beam with wavelength $\lambda = 1.064\,\mu\mathrm{m}$ propagates through a three-layer stratified medium before reaching the trapping region. The stratified medium (as described in the experimental section) consists of an immersion oil layer with refractive index $1.516$, a $200\,\mu\mathrm{m}$-thick coverslip with refractive index $1.82$, and a sample chamber containing dispersed particles in water with refractive index $1.33$. With respect to the nominal focal position of the objective, the oil--coverslip and coverslip--sample interfaces are located at axial positions $-190\,\mu\mathrm{m}$ and $10\,\mu\mathrm{m}$, respectively. The laser beam is focused using a $100\times$ oil-immersion objective with numerical aperture $\mathrm{NA}=1.41$, focal length $f=1.8\,\mathrm{mm}$, and back-aperture radius $w=1.67\,\mathrm{mm}$. The incident beam is assumed to be an $x$-polarised Gaussian beam with waist radius $8.37\,\mathrm{mm}$ before entering the objective. The trapping particles are spherical dielectric polystyrene beads with refractive index $1.59$. The parameters used in the numerical simulations are chosen to match the experimental conditions as closely as possible, ensuring a direct and meaningful comparison between the theoretical predictions and experimental observations. Unless stated otherwise, all distances reported in the simulations are measured relative to the nominal focal position of the beam. The calculated optical forces are reported in arbitrary units; however, the relative magnitudes across different plots are directly comparable. Since the optical force scales linearly with optical power in the present regime, the reported force profiles faithfully capture the relative trends, independent of the absolute power normalization.

Before investigating the influence of the RI mismatch on the optical trapping scenario, it is useful to discuss the distribution of the three-dimensional optical force under RI-matched conditions, serving as a reference case for understanding the modifications introduced by spherical aberration in the experimentally relevant mismatched configuration. The origin is situated at the geometrical focus of the lens. In our simulation, we assume the interface between Oil + coverslip and water to be located at $z = - 5 \;\mu m$ (as depicted in figure \ref{fig:coordinate}). We then proceed to plot the radial and axial forces for a trapped polystyrene particle (radius, $a= 1.5 \;\mu m$) in Fig.~\ref{fig:optical force}(a and b) using the simulation parameters described above. Now, the radial distribution in the longitudinal plane, $F_x(x,z)$, suggests that it is only possible to trap the particle along the z-axis.  The axial equilibrium condition $(F_z = 0)$ forms a semicircular contour in the (xz)-plane (as shown in black dashed lines in Figs.~\ref{fig:optical force}(a,b)), corresponding to the locus where the axial gradient force balances the radiation pressure force. In our analysis, a stable optical trap is identified at the intersection of the three equilibrium manifolds satisfying $(F_x = F_y = F_z = 0)$. Using this criterion, a stable trapping position for a polystyrene particle of diameter $(3  \mu\mathrm{m})$ is only obtained at $((x,y,z) = (0,\,0,\,-1.44 ~\mu\mathrm{m}))$.

Now -- for the mismatched condition, experimentally, both on-axis and off-axis trapping are observed simultaneously at two distinct axial planes, as shown in Fig.~\ref{expFigure}(e). To validate these observations theoretically, we simulate the optical forces in a stratified medium consisting of a $200\,\mu\mathrm{m}$-thick coverslip with RI $n=1.82$, the other RI values remaining the same. 
The oil--coverslip and coverslip--sample interfaces are located at axial positions $z=-180\,\mu\mathrm{m}$ and $z=20\,\mu\mathrm{m}$, respectively, relative to the nominal focal plane of the objective. We then proceed to match our experimental results  for the RI-mismatched condition with those obtained by simulations in the following manner:

\subsubsection*{Validation of transverse extent of trapping region}
Figs.~\ref{fig:optical force}(c,d) reveal that refractive-index mismatch substantially modifies the geometry of the zero-force contours, both axial and radial ($F_z=F_x=0$) in the $xz$-plane, producing an extended equilibrium that spans approximately $4$--$5~\mu\mathrm{m}$ in the transverse direction. Consequently, the aberrated system supports multiple off-axis trapping regions in addition to the conventional on-axis trap. These theoretical predictions are consistent with the experimental observations, where stable off-axis trapping is observed at a radial distance of approx $ 5~\mu\mathrm{m}$ from the center. Thus, optical aberrations effectively reshape the trapping landscape, creating stable trapping regions that are absent in the RI-matched case. This enhanced trapping versatility, however, is accompanied by a reduction in trap stiffness, as indicated by the lower maximum restoring force along the trapping direction compared with the matched configuration. Furthermore, the longitudinal force map exhibits a characteristic conical equilibrium manifold together with the central stable branch, a feature that is entirely absent in the matched case.

\subsubsection*{Validation of the axial separation of trapping planes and the existence of repulsive forces}
The axial force maps ($F_z$) and radial force maps ($F_x$), shown in Fig.~\ref{fig:optical force}(c,d), together reveal the existence of two distinct trapping planes under refractive-index mismatched conditions. Trapping positions correspond to locations where both the axial and radial restoring forces simultaneously vanish, i.e., at the intersections of the $F_z=0$ and $F_x=0$ contours. As evident from Figs.~\ref{fig:optical force}(c,d), these zero-force contours, indicated by black dashed lines, intersect at two spatially separated axial locations. On close inspection of the two figures and as indicated by the values indicated in the respective figure axes, the first intersection occurs at the optical axis ($x=0$) near $z\approx35~\mu$m (red circles in Figs.~\ref{fig:optical force}(c,d)), corresponding to the conventional on-axis trapping plane at the focal region. A second intersection appears approximately $2~\mu$m above this plane, at $z\approx37~\mu$m and $x\approx2~\mu$m (golden circles in Figs.~\ref{fig:optical force}(c,d)).

To assess the three-dimensional stability of these trapping sites, we further examine the radial force distributions in the $xy$-plane at two different axial positions. At the aberrated focal plane ($z=35.111 \; \mu\mathrm{m}$), the radial force field, shown in Fig.~\ref{fig:optical force}(e), behaves conventionally, with stable equilibrium existing only at the optical axis. However, only $2 ~\mu\mathrm{m}$ above this plane, the force landscape changes qualitatively (Fig ~\ref{fig:optical force}(f)), giving rise to stable off-axis equilibrium points that agree well with the experimentally observed trapping locations. This prediction is further supported by the time-lapse images in Fig.~\ref{fig:timesnap}(a-d), where a particle approaching the optical axis is deflected away from the center instead of being trapped. The corresponding radial force maps reveal a repulsive annular region surrounding the optical axis (shown in violet in Fig. ~\ref{fig:optical force}(f)), which acts as a potential barrier and redirects particles into the aberration-induced off-axis trapping sites. The agreement between the experimentally observed trapping positions and the calculated restoring-force distributions therefore establishes a direct quantitative correspondence between theory and experiment.
\begin{figure*}[t]
    \centering
    \includegraphics[width=1\textwidth]{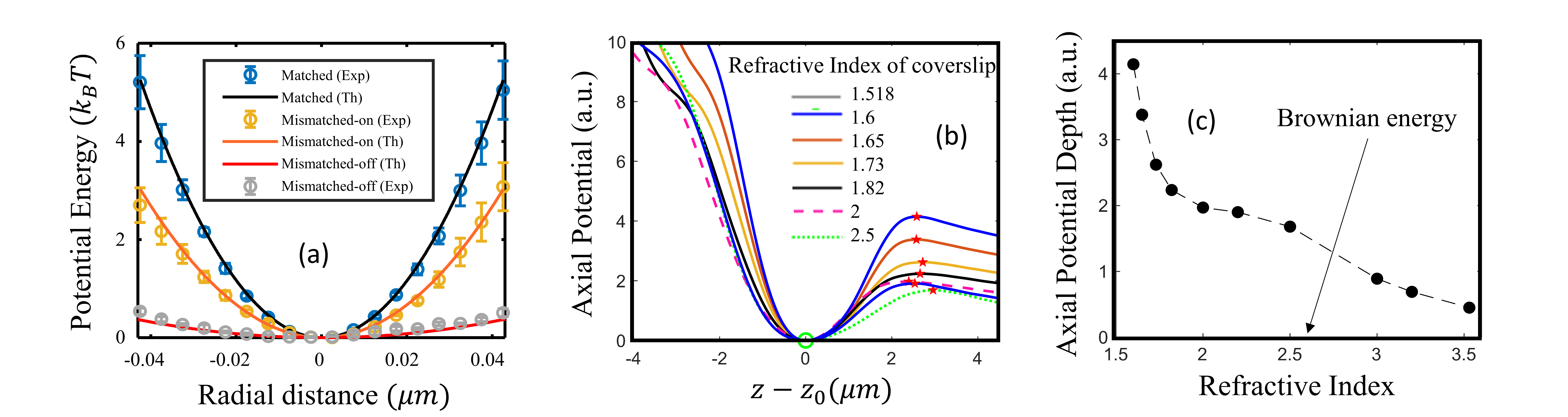}
   \caption{Optical potential.a) Comparison of the theoretically calculated and experimentally measured optical potentials for different trapping configurations. (b) Axial optical potential as a function of the relative displacement from the stable axial equilibrium position for coverslips with different refractive indices. Here, $z_0$ denotes the equilibrium position of the trapped particle for each coverslip. (c) Variation of the axial potential depth with the refractive index of the coverslip.}
    \label{fig:optical potential}
\end{figure*}

\subsubsection*{Validation of trapping strength from the potential depth}
Although the force distributions identify the locations of stable equilibrium, they do not quantify the strength of confinement. This information is provided by the optical potential, which describes the energy landscape experienced by a trapped particle. The optical potential is obtained from the force field, where deeper potential wells correspond to stronger confinement and higher energy barriers to thermally driven escape. Unlike force maps, the optical potential can also be experimentally reconstructed from the Brownian motion of a trapped particle, providing a direct means to validate the theoretical model. To experimentally determine the potential, the particle trajectory is recorded using a high-speed camera at $1000 \mathrm{Hz}$, which is well above the corner frequencies of the optical traps considered here. Assuming thermal equilibrium, the particle position obeys the Maxwell--Boltzmann distribution,

\begin{align}
\rho(x)=\rho_0\exp\left(-\frac{U(x)}{k_BT}\right),
\end{align}

yielding the optical potential

\begin{align}
U(x)=-k_BT\ln[\rho(x)]+U_0,
\label{eq:experimental_potential}
\end{align}

where $U_0$ is an arbitrary energy offset. The same experimental datasets used in the power spectral density analysis are employed to reconstruct the trapping potential \cite{JonesOpticalTweezers2015}.

For comparison, the theoretical potential is obtained directly from the simulated force profile,

\begin{align}
U(x)=-\int_{x_{\mathrm{ref}}}^{x}F(x',z_0),dx',
\end{align}

where $x_{\mathrm{ref}}$ is chosen sufficiently far from the trapping region such that the optical force is negligible. Since the simulated potential is defined up to an arbitrary scaling constant, the theoretical profile is first calibrated using the matched configuration. The resulting scaling factor is subsequently applied to all mismatched cases without modification. As shown in Fig.~\ref{fig:optical potential}(a), the reconstructed experimental potentials are in good agreement with the theoretical predictions, confirming that the simulated force distributions accurately capture the trapping landscape. In agreement with the force analysis, the mismatched configuration exhibits a noticeably shallower potential well, reflecting weaker optical confinement.

Having established the agreement between theory and experiment, we next examine how RI mismatch influences the axial trapping potential. For each coverslip refractive index, the equilibrium position $z_0$ is first determined from the condition $F_z=0$, while the interface locations are fixed at $z_1=-205~\mu\mathrm{m}$ and $z_2=-5~ \mu\mathrm{m}$. The axial potential is then calculated by integrating the corresponding axial force, taking the potential to be zero in the far-field region. To facilitate comparison, all potential profiles are plotted as functions of the relative displacement $(z-z_0)$ so that their minima coincide at the origin. Figure~\ref{fig:optical potential}(b and c) reveals a systematic reduction in potential depth with increasing refractive-index mismatch. This behavior arises from the progressive increase in spherical aberration, which broadens the focal field and weakens the intensity gradients responsible for optical confinement. Consequently, the restoring force decreases, producing a broader and shallower potential well. As the mismatch becomes sufficiently large, the trap depth approaches the thermal energy scale (approximately $7 k_BT$), beyond which Brownian fluctuations overcome the optical restoring force and stable trapping is no longer maintained.

Note that in the present theoretical framework, interparticle interactions arising from optical binding have not been incorporated. Nevertheless, the model validates the experimentally observed trapping behavior with good agreement. This close agreement suggests that the contribution of optical binding in the transverse direction is limited under the present experimental conditions, and that the observed dynamics are predominantly governed by the trapping forces included in the model. A comprehensive treatment incorporating interparticle optical-binding interactions will be presented in future theoretical studies.

\section{Conclusion}

In this work, we have demonstrated a unified theoretical, numerical, and experimental framework to achieve extended volumetric optical trapping in a stratified medium using a single Gaussian beam. By systematically engineering the refractive index gradient across the trapping medium, we show that the focal field undergoes controlled axial elongation accompanied by the formation of multiple localized intensity maxima. These features enable stable trapping of particles at distinct axial planes, including both radially on- and off-axis equilibrium positions, thereby realizing a volumetric trapping landscape without the need for holographic beam shaping. Experimentally, we demonstrate such robust 3D trapping using high contrast RI-stratified media, and proceed to develop an external imaging system capable of clearly resolving out particles trapped at different axial and radial positions. We then go on to calibrate the trapping volume spatially with high accuracy and precision. Further, we validate our experimental results by employing a theoretical formalism based on the Debye–Wolf diffraction formalism combined with Generalized Lorenz–Mie Theory, that accurately captures the electromagnetic field redistribution arising from RI stratified media. Importantly, we establish that a reduced three-layer stratified model is sufficient to describe the dominant physics governing transverse trapping, as the contributions from weak reflections at additional interfaces primarily affect the axial intensity modulation. Our numerical simulations further reveal that RI mismatch plays a critical role in restructuring the optical force landscape through spherical aberration. In contrast to the tightly confined equilibrium region in the RI–matched case, mismatched conditions lead to a significant broadening of stable trapping regions and the emergence of off-axis equilibrium sites. The computed force maps and potential landscapes quantitatively reproduce the experimentally observed trapping positions, establishing a direct correspondence between aberration-induced field redistribution and particle confinement. Moreover, the analysis of axial potentials shows a monotonic reduction in trap depth with increasing refractive-index mismatch, highlighting the trade-off between multi-site trapping capability and trapping strength. In general, this study introduces a simple yet powerful strategy for three-dimensional optical manipulation using RI engineering in stratified media. The ability to control the spatial distribution, number, and stability of trapping sites using a single Gaussian beam opens new possibilities for scalable optical assembly, complex colloidal structuring, and advanced biophotonics applications. Indeed, even beyond optical trapping, the insights presented here actually provide a broader framework for tailoring light–matter interactions in inhomogeneous optical environments, which can be studied more rigorously in future work.

\section*{Acknowledgements}
Sramana Das and Suvajit Dey contributed equally to this paper. The authors acknowledge Indian Institute of Science Education and Research Kolkata for infrastructure and funding support. SD acknowledges the University Grants Commission, Ministry of Education for research fellowship. 

\bibliographystyle{unsrt}

\end{document}